\documentclass[aps,prd,preprint,superscriptaddress,tightenlines,nofootinbib,showpacs]{revtex4}

\usepackage{graphicx}
\usepackage{dcolumn}
\usepackage{bm}

\begin{document}

\preprint{CLNS 08/2028}       
\preprint{CLEO 08-11}         
\def\etaP{\eta^{\prime}}

\title{Measurement of
Exclusive Baryon-Antibaryon Decays of $\chi_{cJ}$ Mesons}

\author{P.~Naik}
\author{J.~Rademacker}
\affiliation{University of Bristol, Bristol BS8 1TL, UK}
\author{D.~M.~Asner}
\author{K.~W.~Edwards}
\author{J.~Reed}
\affiliation{Carleton University, Ottawa, Ontario, Canada K1S 5B6}
\author{R.~A.~Briere}
\author{T.~Ferguson}
\author{G.~Tatishvili}
\author{H.~Vogel}
\author{M.~E.~Watkins}
\affiliation{Carnegie Mellon University, Pittsburgh, Pennsylvania 15213, USA}
% ...
\collaboration{CLEO Collaboration}
\noaffiliation

\date{June 10, 2008}

\begin{abstract}
Using a sample of $2.59 \times 10^7$ $\psi(2S)$ decays collected
by the CLEO--c detector, we present
results of a study of $\chi_{cJ}$ ($J$=0,1,2) decays
into baryon-antibaryon final states. 
We present the world's most precise measurements 
of the $\chi_{cJ}\to p\overline{p}$ and $\chi_{cJ}\to \Lambda \overline{\Lambda}$
branching fractions,
and the first measurements of $\chi_{c0}$ decays to other hyperons.
These results illuminate
the decay mechanism of the $\chi_c$ states.

\end{abstract}

\maketitle


In the standard quark model, the $\chi_{cJ}$ ($J=0,1,2$), 
mesons are $c\bar{c}$ states
in an $L=1$ configuration. The $\chi_{cJ}$ mesons 
are not produced directly
in $e^+e^-$ annihilations. However, the large branching fractions
of $\psi(2S)\to\chi_{cJ}\gamma$ make $e^+e^-$ collisions
at the $\psi(2S)$ energy a very clean  
environment for $\chi_{cJ}$ investigation. 

The available data on the decays of the $\chi_{cJ}$ mesons into baryon-antibaryon
pairs has so far been very limited. The easiest of these final states to detect and measure
is $p\overline{p}$ \cite{PDG}.
The partial width for $\chi_{c0} \to p\overline{p}$ was originally predicted to 
be zero in some models
due to the Helicity Selection Rule \cite{BRODSKI}. However, this rule 
has long been known to 
be strongly violated. More recent work has concentrated on the importance of
the Color Octet Mechanism (COM), 
which treats the $\chi_c$ states as more than just pure $q\overline{q}$
states and incorporates octet operators in the transition 
matrix elements to a given final state in order 
to calculate two-body exclusive decay rates \cite{COM}. 
In particular, Wong \cite{WONG} used the COM to explain the high rate of $\chi_{cJ}\to p \overline{p}$
and made predictions for $\chi_{cJ}\to \Lambda \overline{\Lambda}$. However, these predictions
fell well below the 
low-statistics measurements from BES \cite{BESLambda} that imply 
${\mathcal{B}}(\chi_c\to\Lambda\overline{\Lambda})/{\mathcal{B}}(\chi_c\to p\overline{p}) \approx$ 2 to 4 for all three $\chi_c$ 
states. 
It has since been postulated that such large ratios can be explained without
using the COM, and, instead including a more detailed quark model of the daughter products \cite{PING}.
However, the resulting
predictions depend greatly on the details of this model, and it is clear that more
experimental input is needed. 
In this paper, we analyze a large sample of $\psi(2S)$ decays 
and present results on 
two-body decays of the $\chi_{cJ}$ mesons into $p\overline{p}$, 
$\Lambda\overline{\Lambda}$, 
$\Sigma^0\overline{\Sigma}^0$,
$\Sigma^+\overline{\Sigma^+}$,
$\Xi^-\overline{\Xi^-}$,
and
$\Xi^0\overline{\Xi}^0$. 

The data were taken by the CLEO-c detector \cite{CLEOC} 
operating at the Cornell 
Electron Storage Ring with $e^+e^-$ collisions at 
a center of mass energy corresponding to the $\psi(2S)$
mass of 3.686 GeV/$c^2$. The data correspond
to an integrated luminosity of  
56.3 ${\rm pb^{-1}}$, and
the total number of $\psi(2S)$ events is calculated as $2.59\times 10^7$,
determined according to the
method described in \cite{PSI2S}.

Photons were detected using the 
CsI crystal calorimeter \cite{CLEOII}, which has an energy resolution of
2.2\% at 1 GeV, and 5\% at 100 MeV.  
To discriminate protons from kaons and pions, we combined specific ionizations
$(dE/dx)$ measured in the drift chamber and log-likelihoods obtained
from the ring-imaging \v Cerenkov detector (RICH) \cite{RICH} to form a log-likelihood difference: 
${\cal L}(p-\pi)={\cal L}_{\mathrm RICH}(p)-{\cal L}_{\mathrm RICH}(\pi)+
\sigma^2_{dE/dx}(p)-\sigma^2_{dE/dx}(\pi)$, where negative ${\cal L}(p-\pi)$
implies the particle is more likely to be proton than a pion. For all protons in the 
events we require ${\cal L}(p-\pi)<0$ and ${\cal L}(p-K)<0$. 
This is a very efficient requirement. 

We reconstruct the hyperons in the following decay modes:
$\Lambda \to p\pi^-$ (branching fraction 63.9\%) \cite{PDG} , $\Sigma^+ \to p\pi^0 (51.6\%), \Sigma^0 \to 
\Lambda \gamma (100\%), \Xi^- \to \Lambda \pi^-(99.9\%), $ and $ \Xi^0 \to 
\Lambda \pi^0(99.5\%)$. 
Our hyperon detection follows the technique explained elsewhere \cite{dibaryon}.
Briefly, to reconstruct $\Lambda$ 
candidates, proton candidates are combined with 
charged tracks that are assumed to be pions.  
The $p\pi$ combination is required to be within 
10 MeV of the known $\Lambda$ mass and then is kinematically constrained to that value.
Similarly, $\Xi^-$ candidates are built from these $\Lambda$ candidates with
the addition of another appropriately charged track assumed to be a pion. 
The $\Lambda\pi$ vertex was required to be closer to the beamspot than the $\Lambda$
decay point. The $\Sigma^0$ candidates were formed from the combination of 
$\Lambda$ candidates and a cluster of greater than 50 MeV energy detected in the
crystal calorimeter, not matched to the trajectory of a charged track,
and consistent in shape with that expected from a photon. The $\Sigma^+$ and $\Xi^0$ 
reconstruction is complicated by the fact that 
we cannot use the beamspot for the point of origin
of the photons. 
A kinematic
fit is made to the hypothesis that the parent hyperon originated at the beamspot,
and decayed after a positive path-length at a point taken to be the origin
of the $\pi^0 \to \gamma \gamma$ decay. A requirement was placed on the 
$\chi^2$ of the fit to this topology, which includes the fit to the $\pi^0$
mass from the newly found decay vertex.
In all cases, hyperon candidates within $3 \sigma$ of their nominal masses are 
considered for further analysis, and their four-momenta are then constrained to 
the nominal hyperon mass. 
These kinematic constraints were sufficient to ensure that cross-feed
background from real $\chi_{cJ}$ decays, for instance
$\Lambda\overline{\Lambda}\pi^+\pi^-$ in the $\Xi^-\overline{\Xi^-}$ sample, 
was negligible.

For events with two distinct baryon candidates, 
we combine the candidates into a 
$\chi_{c}$ candidate. At this stage of the analysis, 
the invariant mass resolution of the $\chi_c$ is around 
15 MeV$/c^2$.
We then search for any unused photon in the event 
and add that to the $\chi_c$ candidate to form
a $\psi(2S)$ candidate. This $\psi(2S)$ is then kinematically 
constrained to the four-momentum of the beam, 
the energy of which is calculated using the known
$\psi(2S)$ mass. The momentum is non-zero due to the
finite crossing angle ($\approx 3$ mrad per beam) in CESR. 
To make our final selection, 
we require the $\psi(2S)$ candidate to have a  
$\chi^2$ of less than 25 for the four degrees of freedom for this fit; 
this requirement rejects most background combinations. 
This kinematic fit greatly improves the mass resolution of the 
$\chi_c$ candidate.

To study the efficiency and resolutions, 
we generated Monte Carlo samples
for each $\chi_c$ into each final state 
using a GEANT-based detector simulation \cite{GEANT}.
The simulated events have an angular distribution of
$(1+\alpha \cos^2 \theta)$, where $\theta$ is the 
radiated photon angle relative to the
positron beam direction, and $\alpha=$ 1, -1/3, and 1/13 for the 
$\chi_{c0}$, $\chi_{c1}$, and $\chi_{c2}$ respectively, 
in accordance with expectations for an E1 transition.
The mass resolution and efficiencies are shown in Table~I. 
The resolutions
are approximated
by single Gaussian signal functions. The efficiencies 
shown include all the relevant 
branching fractions. \cite{PDG}
\begin{table}[htb]
\caption{Efficiencies (in \%) obtained
from analysis of Monte Carlo generated events, and yields found in the
data sample.}

\begin{tabular}{l|cc|cc|cc}
\hline
\hline
Mode  & \multicolumn{2}{c|}{ $\chi_{c0}$ } & 
\multicolumn{2}{c|}{ $\chi_{c1}$} &
\multicolumn{2}{c}{ $\chi_{c2}$} \\
\hline
      & Yield & Efficiency(\%) & Yield & Efficiency (\%)& Yield & Efficiency(\%) \\
 \hline
$p \overline{p}$       & $383\pm22$ & $62.4$ & $141\pm13$ & $66.6$ & $121\pm12$ & 65.5  \\
$\Lambda\overline{\Lambda}$ & $131\pm12$ & $16.2$ & $46.0\pm7.2$ & 17.1 & $71.0\pm9.2$&$17.3$   \\
$\Sigma^0\overline{\Sigma}^0$ & $78\pm10$ & 4.1 & $3.8\pm2.5$ & 4.0 & $7.5\pm3.4$& 4.0   \\
$\Sigma^+\overline{\Sigma^+}$ & $39\pm7$ & 5.2 & $4.3\pm2.3$ & 5.0 & $4.0\pm2.3$& 4.7   \\
$\Xi^-\overline{\Xi^-}$ & $95\pm11$ & 7.7 & $16.4\pm4.3$ & 8.2 & $29\pm5$& $8.4$   \\
$\Xi^0\overline{\Xi}^0$  & $23.3\pm4.9$ & 2.9 & $1.7\pm1.4$ & $2.9$& $2.9\pm1.7$ & 2.9   \\
\hline
\hline
\end{tabular}
\end{table}
                                                                                    
The final invariant mass distributions are shown in Fig.~1.
These plots are each fit with three signal shapes comprising Breit-Wigner
functions convolved
with Gaussian resolutions, together with a constant background term. 
The masses and widths
of the Breit-Wigner functions 
were fixed according to the current averages \cite{PDG}, and 
the widths of the Gaussian resolution functions were 
fixed at the values found from Monte Carlo simulation (ranging
from 3.6-5.1 MeV$/c^2$ depending on the spin of the $\chi_c$ and the 
decay mode). The yields from these fits are tabulated in Table~I.

To convert the yields to branching fractions, we divide by the 
product of the number of $\psi(2S)$ events in the data sample,
the detector efficiency,
and the 
branching fractions for $\psi(2S)$ into $\chi_{cJ}$. For the 
last factor 
we use the CLEO measurements of $\mathcal{B}(\psi(2S)\to\gamma\chi_{c0})=
9.22\pm0.11\pm0.46$\%,   
$\mathcal{B}(\psi(2S)\to\gamma\chi_{c1})=
9.07\pm0.11\pm0.54$\%, and
$\mathcal{B}(\psi(2S)\to\gamma\chi_{c2})=
9.33\pm0.14\pm0.61$\% 
\cite{PSI2S}. The results are tabulated in 
Table~II.

We consider systematic uncertainties from 
many different sources.
All modes have a 2\% uncertainty from the total number of $\psi(2S)$ decays \cite{PSI2S}.
The requirement on the $\chi^2$ of the
constraint to the beam four-momentum has been checked by changing 
the cut and noting
the change in the yield in these, and other similar decay modes.
Based on
this study we place a systematic uncertainty of 2.5\% on the efficiency 
of this 
requirement. The uncertainties due to track reconstruction are 0.3\% per charged track. 
The limited Monte Carlo statistics introduces an uncertainty that 
is always a small fraction of the statistical uncertainty in the data.
Using comparison of data and Monte Carlo simulation of hyperon and anti-hyperon
yields from the $\psi(2S)$, we checked our modeling of the hyperon selection
efficiency.
The assigned
systematic uncertainty arising from this study was up to 3\% per hyperon.
The systematic uncertainty due 
to the photon detection and shower-shape criteria 
is set at 2\% per photon.
In the case of the $\chi_{c1}$ decaying into two spin one-half particles,
the two daughters can have their spins either parallel or antiparallel,
and in the $\chi_{c2}$ case there are even more possibilities of combinations
of intrinsic spins and relative angular momentum.
These helicity correlations are not well known in the case of decays
into baryons, and this 
introduces a small uncertainty in the modeling of the efficiencies.
We investigated the effects of 
helicity amplitudes on our efficiency by generating Monte
Carlo with a variety of different helicities
and found small variations. 
From this study, we assign a 1\% uncertainty in the efficiency
of the $\chi_{c1}$ and 2.5\% of the $\chi_{c2}$.
When calculating the final branching
fractions, we add the above systematic uncertainties in quadrature.
The uncertainty due to the
$\psi(2S)\to\gamma\chi_c$ branching fractions is kept separate and quoted
as a second systematic uncertainty.

For evaluating the limits in the cases where there is no
significant signal, we take the probability density function and convolve this 
with Gaussian systematic uncertainties. We then find the  
branching fraction that includes 90\% of the total area.
\begin{table}[htb]
\caption{Branching fraction results (in units of $10^{-5}$) for each decay mode.
The uncertainties are statistical, systematic due to this measurement, and systematic
due to the $\psi(2S)\to\chi_{cJ}\gamma$ rate, respectively. 
The limits on the branching fractions
include all systematic uncertainties.}
\begin{tabular}{lcccc}
\hline
\hline
Mode  &  & $\chi_{c0}$ &  $\chi_{c1}$ & $\chi_{c2}$ \\
\hline
$p \overline{p}$       
 &$\ $This Work$\ $ & $25.7\pm1.5\pm1.5\pm1.3$ & $9.0\pm0.8\pm0.4\pm0.5$ & $7.7\pm0.8\pm0.4\pm0.5$ \\  
 & PDG        & $22.5\pm2.7$  & $7.2\pm1.3$ & $6.8\pm0.7$  \\

$\Lambda\overline{\Lambda}$ 
 &$\ $ This Work$\ $ & $33.8\pm3.6\pm2.2\pm1.7$ & $11.6\pm1.8\pm0.7\pm0.7$ &  $17.0\pm2.2\pm1.1\pm1.1$ \\ 
 & PDG       & $47.0\pm16.0$ & $26.0\pm12.0$ & $34.0\pm17.0$  \\

$\Sigma^0\overline{\Sigma}^0$ 
  &$\ $ This Work$\ $ & $44.1\pm5.6\pm4.2\pm2.2$ & $<4.4$ & $<7.5$ \\
  & PDG      &   &  &   \\

$\Sigma^+\overline{\Sigma^+}$ 
 &$\ $ This Work$\ $ & $32.5\pm5.7\pm4.0\pm1.7$ & $<6.5$ & $<6.7$  \\
 & PDG & & &\\

$\Xi^-\overline{\Xi^-}$ 
 &$\ $ This Work$\ $ & $51.4\pm6.0\pm3.9\pm2.6$ & $8.6\pm2.2\pm0.6\pm0.5$ & $14.5\pm3.0\pm1.2\pm0.9$ \\
 & PDG        & $<103$\footnote{
The BES central value \cite{BESXi} for this measurement is $(53\pm 27\pm 9) \times 10^{-5}$, in good
agreement with this work. 
}
 &  $<34$ & $<37$  \\

$\Xi^0\overline{\Xi}^0$  
 &$\ $ This Work$\ $  & $33.4\pm7.0\pm4.5\pm1.7$  & $<6.0$ & $<10.6$ \\
 & PDG & & & \\

\hline
\hline
\end{tabular}
\end{table}

In summary, 
we measure branching fractions for $\chi_{c0}$ decays into 
$p\overline{p}$, $\Lambda\overline{\Lambda}$, $\Xi^-\overline{\Xi^-}$,
$\Xi^0\overline{\Xi^0}$, $\Sigma^0\overline{\Sigma^0}$ and 
$\Sigma^+\overline{\Sigma^+}$. 
For $\chi_{c1}$ and $\chi_{c2}$
we find significant signals and measure 
branching fractions into the first three of the
above decay modes.
Upper limits on branching fractions are obtained for the 
remainder of the modes.
In the case of $\chi_{cJ}\to p \overline{p}$ and 
$\chi_{cJ} \to \Lambda \overline{\Lambda}$, these measurements
are the most precise to date; in the other modes they 
represent first measurements.
Our values of the branching fractions for $\Lambda\bar{\Lambda}$ are 
well below those reported by BES, but confirm the 
trend that the branching fractions into 
$\Lambda\bar{\Lambda}$ are higher than those for $p\bar{p}$. 
The fact that the $\chi_{c0}$ branching fractions
into $\Sigma\overline{\Sigma}$ and $\Xi\overline{\Xi}$ are all greater than
that of 
$\chi_{c0}\to p\overline{p}$, a trend not mirrored in the 
$\chi_{c1}$ and $\chi_{c2}$ decays, is not in agreement with
naive expectations for the decay of an SU(3) singlet.

We gratefully acknowledge the effort of the CESR staff
in providing us with excellent luminosity and running conditions.
D.~Cronin-Hennessy and A.~Ryd thank the A.P.~Sloan Foundation.
This work was supported by the National Science Foundation,
the U.S. Department of Energy,
the Natural Sciences and Engineering Research Council of Canada, and
the U.K. Science and Technology Facilities Council.

\begin{figure}[htb]

\includegraphics*[width=2.5in]{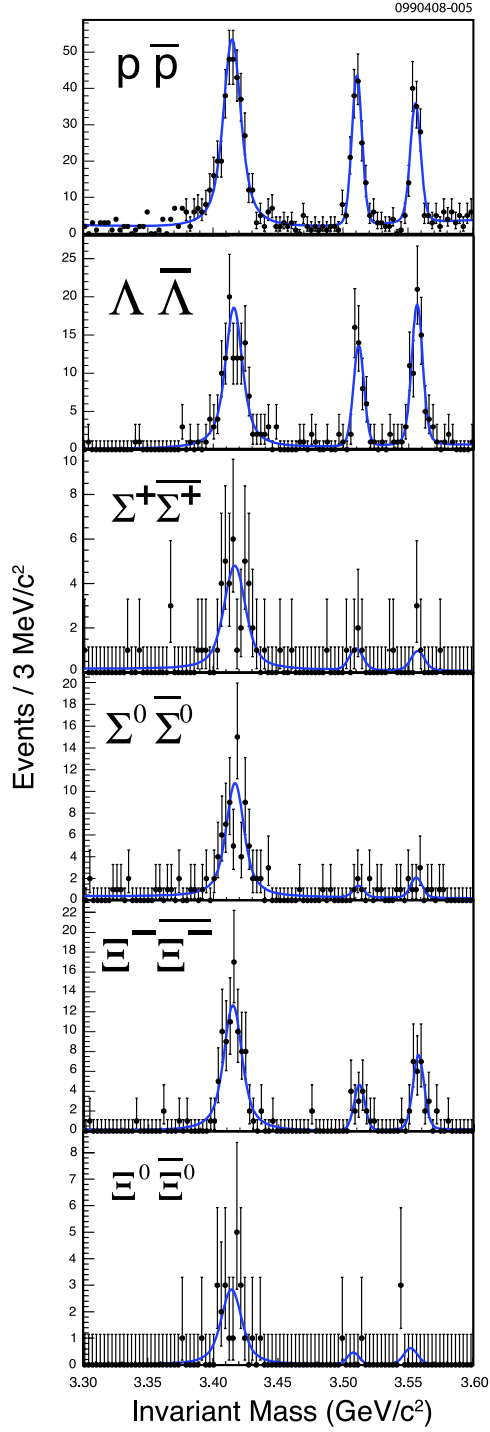}
\caption{Invariant mass distributions for $p\overline{p}$, $\Lambda\overline{\Lambda}$,
$\Sigma^0\overline{\Sigma}^0$, $\Sigma^+\overline{\Sigma^+}$, $\Xi^-\overline{\Xi^-}$,
$\Xi^0\overline{\Xi}^0$.  
The fits are described in the text.
}

\end{figure}

\end{document}